\documentclass[fleqn,twoside]{article}
\usepackage{espcrc2}

\readRCS
$Id: espcrc2.tex,v 1.2 2004/02/24 11:22:11 spepping Exp $
\newcommand{\AmS}{{\protect\the\textfont2
  A\kern-.1667em\lower.5ex\hbox{M}\kern-.125emS}}

\title{
{\small 
\begin{flushleft}
DESY 04--115 \hfill {\tt hep-ph/0407045}\\
        June 2004 \hfill SFB/CPP--04--24 \end{flushleft}} 
NNLO coefficient functions of Higgs and Drell--Yan cross sections
in Mellin space}

\author{J. Bl\"umlein\address[DESY]{Deutsches Elektronen Synchrotron, 
        DESY, Platanenallee 6,
        D-15738 Zeuthen, Germany.}
and
        V. Ravindran\addressmark$^,$\address[HRI]{Harish-Chandra Research
        Institute,
        Chhatnag Road, Jhunsi, Allahabad, India.}
         \thanks{Based on the talk given at
                   7th DESY Workshop on Elementary Particle Theory,
                   Loops and Legs in Quantum Field Theory, April 25 -30, 2004,
                   Zinnowitz (Usedom Island), Germany }}

\begin{document}

\begin{abstract}
\noindent
We calculate the Mellin moments of next-to-next-to-leading order
coefficient functions of the Drell-Yan and Higgs production cross
sections. The results can be expressed in term of finite harmonic 
sums which are maximally threefold up to weight four. Various 
algebraic relations among these finite sums reduce the complexity 
of the results suitable for fast numerical evaluations. It is shown
that only five non--trivial functions occur besides Euler's $\psi$--function
in the representation of these Wilson coefficients.
\vspace{1pc}
\end{abstract}

\maketitle

\section{Introduction}
\noindent
With the present $pp$--collider Tevatron \cite{TEV} and the upcoming
large hadron collider~(LHC)~\cite{LHC} at CERN, the need for precise
predictions from theory have become more and more important.
The Drell-Yan~(DY) cross section which is known upto the NNLO level
\cite{DY} not only tests the reliability of perturbative QCD~(pQCD) but
also reduces the uncertainties coming from theory in order to make
background studies more reliable for new particle searches and physics
beyond the Standard Model.  Similarly, higher order corrections to the
total cross section \cite{HGS} for the Higgs boson production make the
predictions more reliable for phenomenological studies. In the following
we will study the structure of these corrections in the Mellin space using
various algebraic identities relating the resulting finite harmonic sums.
This representation allows a considerable reduction of the set of basic 
functions needed to represent the 2--loop Wilson coefficients.

\section{Coefficient Functions}
\noindent
Due to mass factorization, hadronic cross sections such as
for the DY--process and Higgs--boson production can be expressed in terms
of Mellin convolutions
of the perturbatively computable coefficient functions
$\Delta_{ab}(x,Q^2,\mu^2)$ and
non--perturbative parton distributions, $f_a(x,\mu^2)$, of incoming
hadrons.
\begin{eqnarray}
\sigma(x,Q^2)&\!\!\!=\!\!\!&\int_x^1 \!\!{dx_1 \over x_1}\! \int_{x\over x_1}
\!\!{dx_2\over x_2}
f_a(x_1,\mu^2) f_b(x_2,\mu^2)
\nonumber \\
&&\times \Delta_{ab}\Bigg({x \over x_1 x_2},
{Q^2 \over \mu^2}\Bigg)~.
\end{eqnarray}
Though the parton densities are not calculable in the pQCD,
their evolution with respect to a scale is computable,
thanks to the renormalization group~(RG) equations governing mass
factorization.
Hence higher order corrections to hadronic reactions enter
through two sources viz, the coefficient functions and
the RG--equations of parton distribution functions.
In this paper we will concentrate only on the coefficient functions.
Details of the calculation will be given in \cite{JBVR1}.

The perturbatively calculable coefficient functions
are usually computed in terms of the scaling variable $x=Q^2/s$ where
$s$ is the center of mass energy of the incoming partonic system.
$Q^2$ is the invariant mass of the final di--lepton pair (for DY) or
mass squared of the Higgs boson.
The results are expressible in terms of polynomials in $x$, logarithms and
Nielsen integrals \cite{nielsen} defined by
\small {
\begin{equation}
S_{n,p}(x)\!\!=\!\!{(-1)^{n+p-1} \over (n-1)!p!}
\int_0^1 {dz \over z}\! \log^{n-1}(z)\!
\log^p(1-zx)
\end{equation}
}
\normalsize

\noindent
In practice, using  (1), one performs the integration over
$x_1$ and $x_2$ after folding the perturbatively computed
coefficient functions with the appropriate parton distributions.  
This involves further evaluation of various Nielsen integrals given in (2)
or even more general functions. Due to this the complexity
of the numerical evaluation of the hadronic
cross sections is rather large.  In the following,
we will present an alternative treatment of the evaluation
of the total cross section upto NNLO level by working in Mellin space.  Such
techniques have been used in the past to compute
deep inelastic scattering cross sections and indeed they are
found to be most suitable for various resummation programs \cite{JV}.
Recently, similar work has also been performed for the Wilson coefficients
for polarized and unpolarized deeply inelastic
scattering~\cite{JBSM,JB04}.

%

\section{Mellin Moments and Finite Harmonic Sums}
\noindent
Using the Mellin transform \cite{mellin} of a given function $F(x)$ 
\begin{equation}
M\big[F\big](N)=\int_0^1 dx x^{N-1} F(x)
\end{equation}
the cross section (1) in the Mellin space~($N$-space) becomes
\begin{eqnarray}
M\big[\sigma\big](N,Q^2)&=&M\big[f_a\big](N,\mu^2) M\big[f_b\big](N,\mu^2)
\nonumber\\[2ex]
&&\times M\big[\Delta_{ab}\big]\Big(N,{Q^2 \over \mu^2}\Big)~.
\end{eqnarray}
As is clear from the above equation, the convolutions of
functions in $x$-space reduce to simple products of
Mellin moments in $N$-space.  The Mellin moments of these functions
can be analytically continued \cite{JB3} to complex values of $N$
so that one can use various analyticity properties of these functions in
the complex $N$-plane to evaluate them efficiently.

For our analysis, the starting point is eqn.~(1) with given
parton densities $f_a(x,\mu^2)$ and known
coefficient functions $\Delta_{ab}(x,Q^2)$ computed upto NNLO in pQCD.
We then compute the Mellin moments of these functions in $N$-space
and perform their analytic continuation to complex values of $N$.
At the end,
we substitute them back into eqn.~(4) and
perform the inverse Mellin transformation to arrive at
the results in $x$-space using a suitable contour in the complex
$N$-plane. Compared to the direct numerical convolution in $x$--space
the numerical computation using the $N$--space results is
much faster.

Since we are dealing with the DY/Higgs total production  cross
sections, they depend only on two variables, the scaling variable $x$ and
the virtuality $Q^2$, the invariant mass of the final state. 
The set of functions contributing to the two--loop coefficient
functions in $x$ space contains about 80 elements. Their representation
in Mellin space has been given in \cite{JK} in terms of a specific class of 
sums \cite{JK,JVE,JB1,JB2}, the finite harmonic sums.

The finite harmonic sum of $m$-indices is defined as
{\small
\begin{eqnarray}
S_{k_1...k_m}(N)&=&\sum_{n_1=1}^N {[sign(k_1)]^{n_1} \over n_1^{|k_1|}}
                 \sum_{n_2=1}^{n_1} {[sign(k_2)]^{n_2} \over n_2^{|k_2|}}
\nonumber\\[2ex]
&&\cdot \cdot \cdot\sum_{n_m=1}^{n_m-1} 
{[sign(k_m)]^{n_m} \over n_m^{|k_m|}}~,
\end{eqnarray}
} \normalsize
with $l,k_l\not=0$.  For example,
\begin{eqnarray}
&&M\left[\left({\log^3(1-x)\over (1-x)}\right)_+\right](N)=
{1\over 4} S_1^4(N-1)
\nonumber\\[2ex]
&&        +{3\over 2} S_1^2(N-1) S_2(N-1)
        +{3\over 4} S_2^2(N-1)
\nonumber\\[2ex]
&&        +2 S_1(N-1) S_3(N-1)
        +{3\over 2} S_4(N-1)~.
\end{eqnarray}
Here the right hand side is a polynomial out of only single harmonic
sums.
Similarly, a combination of various Nielsen integrals
may have a simple structure in $N$-space \cite{JK}~:
\small
\begin{eqnarray}
&&M\Bigg[S_{12}(-x) +{1\over 2} \Big(2 {\cal L}i_2(-x) \log(1+x)
+\log(x)
\nonumber\\[2ex]
&&
\times \log^2(1+x)\Big)\Bigg](N)
={1 \over N} \Big(-{1 \over 2} \zeta_2 \log(2)
+ {1 \over 8} \zeta_3
\nonumber\\[2ex]
&&
-(-1)^N \big(S_{1,-2}(N)
   +{1\over 2} \zeta_2 \big(S_1(N)-S_{-1}(N)\big)
\nonumber\\[2ex]
&&   +{1 \over 8} \zeta_3-{1 \over 2} \zeta_2 \log(2)\big)\Big)
\end{eqnarray}
\normalsize
The single harmonic sums~($m=1$) obey the following integral
representation
\small
\begin{eqnarray}
S_{\pm k}(N)&=\!\!&\!\!\int_0^1{dx_1\over x_1} \cdot \cdot \cdot \int_0^{x_{k-1}}
{(\pm x_k)^N-1\over x_k \mp 1}
\\[2ex]
&\!\!\!=\!\!\!&{(-1)^{k-1} \over (k-1)!}\int_0^1 dx \log^{k-1}(x)
{(\pm x)^N-1 \over x \mp 1}~. \nonumber
\end{eqnarray}
Using the above representation for the single sums
and the following formula,
\begin{eqnarray}
\sum_{k=1}^N{(\pm x)^k \over k^l}&\!\!\!=\!\!\!&
{(-1)^{l-1} \over (l-1)!}\!\int_0^x \!dz \!\log^{l-1}(z)
{(\pm z)^N-1 \over z \mp 1}
\nonumber\\
\end{eqnarray}
\normalsize
one derives integral representations for the multiple sums
$S_{k_1...k_m}(N)$.  For our purpose, we have made extensive
use of the results given in \cite{JK}.

\section{Algebraic Relations}
\noindent
Finite harmonic sums are related to each other by
various algebraic relations \cite{euler,NIELSHB,JK}.  These relations
result from the shuffle product of two harmonic sums.
Special cases follow from partial or complete index permutations.

The permutation relation for sums with two indices is given by
\begin{eqnarray}
S_{m,n}(N) + S_{n,m}(N) &=& S_m(N) S_n(N) \nonumber\\
& &+ S_{m \wedge n}(N)~, 
\end{eqnarray}
\normalsize
with $m \wedge n = {\rm sign}\{m\} {\rm sign}\{n\}(|m|+|n|)$.

For harmonic sums with three indices 
the following identity is used to derive various relations
\begin{eqnarray}
&&\sum_{perm\{l,m,n\}} S_{l,m,n}(N)=S_l(N) S_m(N) S_n(N)
\nonumber \\&&
+\sum_{inv~~perm\{l,m,n\}}S_l(N)S_{m\wedge n}
+2 S_{l\wedge m\wedge n}
\end{eqnarray}
where "inv perm" means the invariant permutations and "perm"
means permutations.

\begin{eqnarray}
 S_{1,2,1}(N)&=&-2S_{2,1,1}(N)+S_{3,1}(N)
\nonumber\\[2ex]&&
       +S_{1}(N)S_{2,1}(N)+S_{2,2}(N)
\nonumber\\[2ex]
 S_{1,1,2}(N)&=&S_{2,1,1}(N)
           +{1 \over 2}(S_{1}(N)(S_{1,2}(N)
\nonumber\\[2ex]&&
-S_{2,1}(N))+S_{1,3}(N)-S_{3,1}(N))
\nonumber\\[2ex]
 S_{1,-2,1}(N)&=&-2S_{-2,1,1}(N)+S_{-3,1}(N)
\nonumber\\[2ex]&&
     +S_{1}(N)S_{-2,1}(N)+S_{-2,2}(N)
\nonumber\\[2ex]
 S_{1,1,-2}(N)&=&S_{-2,1,1}(N)+S_{-2}(N)S_{2}(N)
\nonumber\\[2ex]&&
    -S_{-2,2}(N)-S_{-2}(N)S_{1,1}(N)
\nonumber\\[2ex]&&
            +S_{1}(N)S_{1,-2}(N)+S_{1,-3}(N)
\nonumber\\[2ex]&&
-S_{1}(N)S_{-3}(N)
\end{eqnarray}

\section{Coefficient Functions in $N$-Space}
\noindent
We have computed the Mellin moments for the DY and the Higgs
coefficient functions using the list given in \cite{JK}
and the algebraic identities given in the previous sections.
For a general investigation of these relations, see \cite{JB1,HOF}. 
The algebraic relations considerably simply
our final result.  At the intermediate stages of
the computation we encounter various complicated sums
such as $S_{1,-1,2}$, $S_{-1,-1,-2}$, $S_{-1,-2,-1}$, $S_{-2,-1,-1}$,
$S_{2,-1,1}$, $S_{1,2,-1}$, $S_{2,1-1}$, $S_{-1,1,2}$, $S_{-1,1,2}$.
At the end, most of these sums disappear
leaving only few sums like $S_{-2,1,1}$, $S_{2,1,1}$.
We have listed below the sums that appear at the end of
the computation.

We obtain eight single sums $S_{\pm i}(N)$ with $i=1...4$,
\small
\begin{eqnarray}
S_{-4}(N)&=&(-1)^{N+1}{1 \over 6}
M\left[{\log^3(x)\over 1+x}\right](N+1)
-{7 \zeta_2^2 \over 20}
\nonumber\\[2ex]
S_{-3}(N)&=&(-1)^N{1 \over 2}
M\left[{\log^2(x)\over 1+x}\right](N+1)
-{3 \over 4} \zeta_3
\nonumber\\[2ex]
S_{-2}(N)&=&(-1)^{N+1}
M\left[{\log(x)\over 1+x}\right](N+1)
-{1 \over 2} \zeta_2
\nonumber\\[2ex]
S_{-1}(N)&=&(-1)^N
M\left[{1\over 1+x}\right](N+1)
-\log(2)
\nonumber\\[2ex]
S_{4}(N)&=&{1 \over 6}
M\left[{\log^3(x)\over 1-x}\right](N+1)
+{2 \over 5} \zeta_2^2
\nonumber\\[2ex]
S_{3}(N)&=&-{1 \over 2}
M\left[{\log^2(x)\over 1-x}\right](N+1)
+\zeta_3
\nonumber\\[2ex]
S_{2}(N)&=&
M\left[{\log(x)\over 1-x}\right](N+1)
+\zeta_2
\nonumber\\[2ex]
S_{1}(N)&=&-
M\left[\left({1\over 1-x}\right)_+\right](N+1)~.
\end{eqnarray}
\normalsize
These harmonic sums can be solely expressed in terms of the Euler
$\psi$--function and the $\beta$--function~\cite{NIELSHB} and their
derivatives, which is related to the former combining two
$\psi$--functions with shifted argument. 
These functions represent at the same time the analytic continuation of
these harmonic sums.
The following five double sums $S_{-3,1}(N)$, $S_{-2,1}(N)$, 
$S_{-2,2}(N)$,
$S_{2,1}(N)$, $S_{3,1}(N)$ occur~:
\small
\begin{eqnarray}
S_{-3,1}(N)&=&(-1)^N
M\left[{{\cal L}i_3(x)\over 1+x}\right](N+1)
\nonumber\\[2ex]&&
+\zeta_2 S_{-2}(N)
-\zeta_3 S_{-1}(N)
-{3 \over 5} \zeta_2^2
\nonumber\\[2ex]&&
+2 {\cal L}i_4\left({1 \over 2}\right)
+{3 \over 4} \zeta_3 \log(2)
\nonumber\\[2ex]&&
-{1 \over 2} \zeta_2 \log^2(2)
+{1 \over 12} \log^4(2)
\nonumber\\[2ex]
S_{-2,1}(N)&=&-(-1)^N
M\left[{{\cal L}i_2(x)\over 1+x}\right](N+1)
\nonumber\\[2ex]&&
+\zeta_2 S_{-1}(N)
-{5 \over 8} \zeta_3
\nonumber\\[2ex]&&
+ \zeta_2 \log(2)
\nonumber\\[2ex]
S_{-2,2}(N)&=&-(-1)^N
M\Bigg[{1\over 1+x}\Big(2 {\cal L}i_3(x)
\nonumber\\[2ex]&&
-\log(x) \Big({\cal L}i_2(x)
+\zeta_2\Big)\Big)\Bigg](N+1)
\nonumber\\[2ex]&&
+\zeta_2 S_{-2}(N)
+2 \zeta_3 S_{-1}(N)
\nonumber\\[2ex]&&
+{71 \over 40} \zeta_2^2
-4 {\cal L}i_4\left({1 \over 2}\right)
\nonumber\\[2ex]&&
-{3 \over 2} \zeta_3 \log(2)
+\zeta_2 \log^2(2)
-{\log^4(2) \over 6}
\nonumber\\[2ex]
S_{2,1}(N)&=& 
M\left[\left({{\cal L}i_2(x)\over 1-x}\right)_+\right](N+1)
+\zeta_2 S_{1}(N)
\nonumber\\[2ex]
S_{3,1}(N)&=& -{1 \over 2}
M\left[{{\cal L}i_2(x)\log(x)\over 1-x}\right](N+1)
\nonumber\\[2ex]&&
+\zeta_2 S_{2}(N)
-{1 \over 4} S_2^2(N)
-{1 \over 4} S_4(N)
\nonumber\\[2ex]&&
-{3 \over 20} \zeta_2^2
\end{eqnarray}
\normalsize
Furthermore, two triple sums $S_{-2,1,1}(N)$, $S_{2,1,1}(N)$ contribute~:
\begin{eqnarray}
S_{-2,1,1}(N)&=&-(-1)^N
M\left[{S_{12}(x)\over 1+x}\right](N+1)
\nonumber\\[2ex]&&
+\zeta_3 S_{-1}(N)
-{\cal L}i_4\left({1 \over 2}\right)
+{1 \over 8} \zeta_2^2
\nonumber\\[2ex]&&
+{1 \over 8} \zeta_3 \log(2)
+{1 \over 4} \zeta_2 \log^2(2)
\nonumber\\[2ex]&&
-{1 \over 24} \log^4(2)
\nonumber\\[2ex]
S_{2,1,1}(N)&=&
M\left[\left({S_{12}(x)\over 1-x}\right)_+\right](N+1)
\nonumber\\ & &
+\zeta_3 S_1(N)
\end{eqnarray}

From our final expression, we observe that only very few functions
do finally contribute  to the Wilson coefficients.
We list them below for completeness:
\begin{eqnarray}
&&{\log^n(x) \over 1-x} \quad \quad \quad \quad \quad
{\log^n(x) \over 1+x}  \quad
n=0,1,2,3
\nonumber\\[2ex]
&&{{\cal L}i_2(x) \log^n(x) \over 1-x} \quad \quad
{{\cal L}i_2(x) \log^n(x) \over 1+x} \quad
n=0,1
\nonumber\\[2ex]
&&{S_{12}(x) \over 1-x} \quad \quad\quad \quad \quad
{S_{12}(x) \over 1+x}  \quad
\nonumber\\[2ex]
&&{{\cal L}i_3(x) \over 1+x}
\end{eqnarray}
Since functions weighted by a factor $\ln^n(x)$ are related to the
un--weighted functions in Mellin space by
\begin{equation}
M[\ln^n(x) f(x)](N) = \frac{\partial^n}{\partial N^n} M[f(x)](N)
\end{equation} 
and single harmonic sums can be expressed by the well--known
$\psi$--functions only five non--trivial functions are needed to express
the two--loop Wilson coefficients for the polarized and unpolarized
DY--process and the scattering cross sections of hadronic Higgs--boson and
pseudoscalar  Higgs--boson production.

The next step is to multiply the appropriate parton densities computed in
complex $N$-space with our coefficient functions
obtained to invert back 
to $x$-space for phenomenological studies.

The Mellin inversion \cite{carlson} of a function $\tilde F(N)$ is given by
\begin{eqnarray}
F(x)={1 \over 2 \pi i} \int_{c-i\infty}^{c+i \infty} dN x^{-N} \tilde F(N)
\end{eqnarray}
Here the parameter $c$ is the intersection of the contour on the real axis
and is chosen in such a way that the integral
$\int_0^1 dx x^{c-1} F(x)$ is convergent.  In other words, $c$ should
lie on the right of rightmost singularity of the function $\tilde{F}(N)$.
The shape of the contour can be deformed at our convenience provided
all singularities are covered.

\section{Conclusion}
\noindent
We have systematically analyzed the mathematical structure
behind the NNLO coefficient functions for DY and Higgs production
using Mellin moment techniques.  Use of various algebraic identities
which relate the finite harmonic sums in Mellin $N$-space
reduces the complexity of the results from around 80 functions to 
only five basic functions, the $\psi$--function and a few derivatives
thereof. This
is very useful not only to understand the nature of higher
order corrections but also to perform fast numerical calculations
at high precision for phenomenological applications and fits to data.
The same structures are found in the case of polarized and
unpolarized 2--loop fragmentation functions~\cite{JBVR2}. Together with
the results of \cite{JBSM} it is now shown that these structures are in 
common for all known massless 2--loop Wilson coefficients.

\vspace{2mm} \noindent
{\bf Acknowledgment} We thank S. Moch for discussions. V.R. would like to
thank DESY for their
kind hospitality extended to him and for a lively meeting, Loops and Legs
2004. This paper was supported in part by DFG Sonderforschungsbereich
Transregio 9, Computergest\"utzte Theoretische Physik and EU grant
HPRN--CT--2000--00149.

\end{document}